\documentclass{article}
\usepackage{spconf,amsmath,graphicx,multirow,float,hyperref}
\usepackage{booktabs}
\usepackage{arydshln}
\usepackage{xcolor}

\newcommand{\ra}[1]{\renewcommand{\arraystretch}{#1}}

\title{Self-supervised learning with bi-label masked speech prediction for streaming multi-talker speech recognition}
\name{\begin{tabular}{c}Zili Huang$^{1\dagger}$, 
Zhuo Chen$^2$, 
Naoyuki Kanda$^2$, 
Jian Wu$^2$, 
Yiming Wang$^2$, 
Jinyu Li$^2$,\\
Takuya Yoshioka$^2$,
Xiaofei Wang$^2$,
Peidong Wang$^2$
\end{tabular}\thanks{$^\dagger$Work performed during an internship at Microsoft.}}
\address{   
$^1$Center for Language and Speech Processing, Johns Hopkins University, Baltimore, MD, USA\\
$^2$Microsoft, One Microsoft Way, Redmond, WA, USA\\}
\begin{document}
\ninept
\maketitle
\begin{abstract}
Self-supervised learning (SSL), which utilizes the input data itself for representation learning, has achieved state-of-the-art results for various downstream speech tasks. However, most of the previous studies focused on offline single-talker applications, with limited investigations in multi-talker cases, especially for streaming scenarios. In this paper, we investigate SSL for streaming multi-talker speech recognition, which generates transcriptions of overlapping speakers in a streaming fashion. We first observe that conventional SSL techniques do not work well on this task due to the poor representation of overlapping speech.  We then propose a novel SSL training objective, referred to as \emph{bi-label masked speech prediction}, which explicitly preserves representations of all speakers in overlapping speech. We investigate various aspects of the proposed system including data configuration and quantizer selection. The proposed SSL setup achieves substantially better word error rates on the LibriSpeechMix dataset.

\end{abstract}
\section{Introduction}
\label{sec:intro}
\vspace{-.5em}


Self-supervised learning (SSL), which extracts supervision signals from data itself, is a fast-growing subcategory of unsupervised learning approaches~\cite{mohamed2022self}. In SSL pipeline, an upstream model is pre-trained on massive unlabeled data with some pretext tasks derived from the data itself. Then it is adapted for specific downstream tasks with a small amount of labeled data by either using the upstream model as a feature extractor~\cite{yang2021superb,chen2022wavlm} or directly fine-tuning it together with additional task-specific layers~\cite{baevski2020wav2vec,hsu2021hubert}.

SSL has been widely explored due to its great performance and low adaptation cost. In speech applications, such SSL-based pre-training has achieved remarkable performance for various downstream tasks including speech recognition~\cite{baevski2020wav2vec,hsu2021hubert}, speaker recognition~\cite{fan2020exploring,chen2022large}, emotion recognition~\cite{pepino2021emotion}, etc. Since the model learns more generalizable task-agnostic representations in the pre-training stage, it only requires a small amount of labeled data in the fine-tuning stage. For example, wav2vec 2.0~\cite{baevski2020wav2vec} outperforms the previous state-of-the-art automatic speech recognition (ASR) results on the LibriSpeech 100h benchmark with just 1h of labeled speech.

Despite the great achievements in various speech tasks, SSL remains under-explored for streaming multi-talker audio processing tasks. It is known that natural human conversations contain a considerable amount of speech overlaps \cite{ccetin2006analysis}, thus handling overlapping speech in real time is in great demand for many real applications. Nevertheless, most of the existing SSL techniques were explored under single-talker speech conditions for both pre-training and fine-tuning~\cite{oord2018representation,riviere2020unsupervised,chung19_interspeech,schneider2019wav2vec,ling2020decoar,baevski2020wav2vec,hsu2021hubert}. Recently, WavLM~\cite{chen2022wavlm} was proposed with the multi-talker data augmentation scheme, called ``utterance mixing'', which was proven to be effective for several multi-talker tasks such as speech separation and speaker diarization \cite{chen2022wavlm}. However, WavLM was designed with an offline model architecture, limiting its usage in streaming scenarios. Moreover, WavLM was pre-trained with a conventional masked speech prediction (MSP) loss, where the model predicts the masked tokens of the primary (i.e. dominant) speaker for augmented multi-talker audios, which could potentially hurt the representations of other speakers. Such training scheme could be sub-optimal for tasks where every speaker are equally important.

In this paper, we investigate SSL-based pre-training for the streaming multi-talker ASR task, in which we perform real-time speech recognition for all speakers in a conversation containing overlapped speech.
We propose a novel \emph{bi-label MSP} objective which forces the model to learn a representation of all speakers in overlapping speech instead of focusing on a single dominant speaker.
We also explore several aspects of the proposed bi-label SSL model to further improve its performance, including the ratio of utterance mixing and the quantizer type. We conducted our experiment based on the streaming multi-talker ASR with the token-level serialized output training (t-SOT) \cite{kanda2022streaming}. Our experimental results on the LibriSpeechMix~\cite{kanda2020serialized} dataset reveal that, with the proposed bi-label MSP objective, appropriate pre-training data configuration, and quantizer, the streaming multi-talker ASR accuracy can be significantly improved.

\begin{figure*}[t]
    \centering
    \includegraphics[width=0.8\linewidth,height=0.24\textheight]{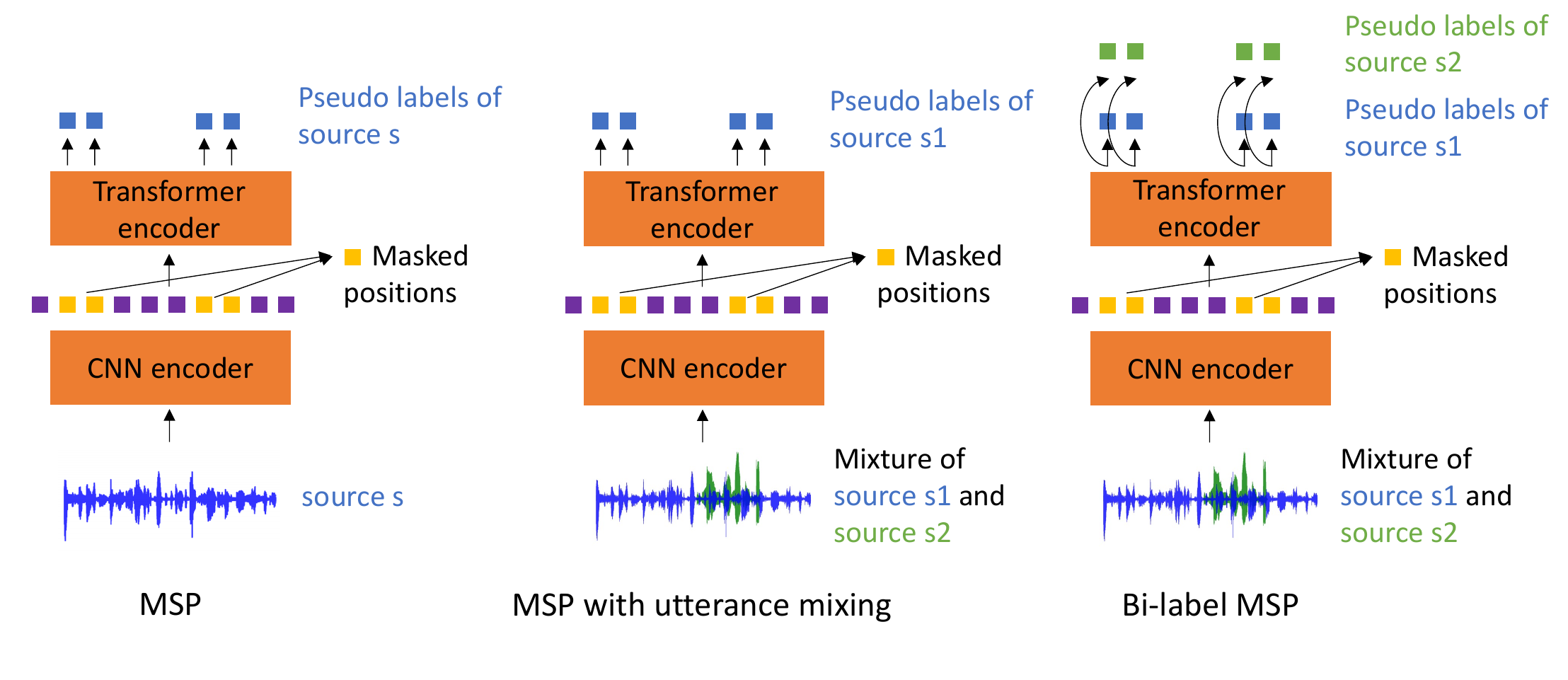}
    \vspace{-9mm}
    \caption{Overview of SSL methods. (Left) MSP, (middle) MSP with utterance mixing, (right) proposed bi-label MSP}
    \label{fig:bi-label}
    \vspace{-5mm}
\end{figure*}

\vspace{-.5em}
\section{Related works}
\label{sec:related work}
\vspace{-.5em}

\subsection{HuBERT and WavLM}
\vspace{-.5em}

Our work is performed under the basis of two SSL models: HuBERT~\cite{hsu2021hubert} and WavLM~\cite{chen2022wavlm}. Similar to BERT~\cite{kenton2019bert}, HuBERT uses MSP as the pretext task: the acoustic embeddings after the convolutional neural network encoder are partially masked, and the transformer encoder is trained to predict the pseudo labels of masked regions (Fig. \ref{fig:bi-label} left). The distribution over the pseudo labels is formulated as
\begin{align}
   p(c|o_t)=\frac{\exp(\cos(o_t \cdot W^P, e_c)/\gamma)} {\sum_{c'=1}^C \exp(\cos(o_t \cdot W^P, e_c')/\gamma)},
\end{align}
where $W^P$ is a projection matrix, $o_t$ is the output logit at the time frame $t$,
$e_c$ is the embedding of the pseudo label $c\in\{1,...,C\}$, $\text{cos}(a,b)$ is the 
cosine similarity between $a$ and $b$, and $\gamma$ is the scale of the logit.
The pseudo labels are generated by clustering either acoustic features (e.g., mel-frequeney cepstrum coefficient or mel-filterbank (FBANK)) or hidden representations from a prior generation of the HuBERT model.

WavLM~\cite{chen2022wavlm} introduced several modifications to HuBERT to enhance spoken content modeling and speaker identity preservation. Firstly, it introduced gated relative position bias \cite{chi2022xlm} instead of convolutional relative position embedding. Secondly, it introduced data augmentation where the input audio was mixed with noise or interfering speech (Fig. \ref{fig:bi-label} middle). Thirdly, the training data size and variety were scaled up to further improve the robustness of learned representations. Among these modifications, WavLM's data augmentation scheme enforces the model to execute a denoising task in addition to the original masked speech prediction task, which significantly improved the performance on speaker-related tasks such as speaker diarization and speech separation.

\vspace{-.5em}
\subsection{Streaming multi-talker ASR based on t-SOT}
\vspace{-.5em}


The token-level serialized output training (t-SOT)~\cite{kanda2022streaming} is a framework to train streaming multi-talker end-to-end (E2E) ASR models \cite{E2EOverview} that can generate transcriptions of multiple overlapping speakers with limited latency. The key of t-SOT lies on the serialization of multi-talker transcriptions. Suppose we have time- and speaker-annotated transcriptions of multiple speakers (e.g., ``hello how are you" from speaker A and ``fine thank you" from speaker B). We first create a single sequence of tokens by simply concatenating the transcriptions of all speakers (e.g., ``hello how are you fine thank you"). We then reorder the tokens in that sequence based on the end time of each token (e.g., ``hello how fine are you thank you"). Finally, we insert a special token {\textlangle}{cc}{\textrangle} when the adjacent tokens are attributed to different speakers (e.g., ``hello how {\textlangle}{cc}{\textrangle} fine {\textlangle}{cc}{\textrangle} are you {\textlangle}{cc}{\textrangle} thank you"). A conventional streaming E2E ASR model, e.g., transformer transducer (TT) \cite{zhang2020transformer,chen2021developing} can be trained with overlapping speech annotated with such serialized transcriptions. In inference, the ASR system generates transcriptions including {\textlangle}{cc}{\textrangle}, which is then ``deserialized'' into two streams of transcriptions based on the estimation of {\textlangle}{cc}{\textrangle}. It was shown that the t-SOT based multi-talker ASR model achieved better accuracy than prior multi-talker models while keeping the model architecture and computational cost the same as conventional single-talker models.




\vspace{-.5em}
\section{Pre-training of Streaming multi-talker ASR with bi-label MSP}
\label{sec:methodology}
\vspace{-.5em}

In this section, we introduce our SSL-based pre-training framework for streaming multi-talker E2E ASR. We first describe our proposed bi-label MSP objective in Section \ref{ssec:bi-label}, and then introduce our strategy to pre-train streaming models in Section \ref{ssec:attention mask}. Finally, we explain data configurations for the pre-training in Section \ref{ssec:other_explore}.



\vspace{-.5em}
\subsection{Bi-label MSP objective}
\label{ssec:bi-label}
\vspace{-.5em}



The original MSP objective is designed to predict pseudo labels of the masked speech region given the surrounding speech as a context. When combined with utterance mixing proposed in WavLM, the MSP enforces the model to learn representations that best estimates the masked speech of the primary (i.e. dominant) speaker while ignoring the speech of the secondary speaker. We speculate that such a formulation may prevent the model from learning a good representation of the speech from the secondary speaker.

In consideration of this, we propose a bi-label MSP objective for pre-training, where the model predicts the pseudo labels of both the primary and the secondary speakers.
The overview of the bi-label MSP is shown on the right side of Fig.~\ref{fig:bi-label}. As shown in the figure, the transformer encoder has two output nodes, one of which predicts the pseudo label of the primary speaker while the other predicts the pseudo label of the secondary speaker. 
If the secondary speaker is not present in the masked regions, a special {\textlangle}{blank}{\textrangle} token is assigned for that region. 
The loss function is formulated as
\begin{align}
    \mathcal{L} = \sum_{t \in M}{-\log(p(r^{pr}_t | o^{pr}_t)) - \log(p(r^{sc}_t | o^{sc}_t))},
\end{align}
where $r^{pr}_t$ and $r^{sc}_t$ are the pseudo labels of the primary and the secondary speaker at time frame $t$,  $o^{pr}_t$ and $o^{sc}_t$ are the output logit for the primary and the secondary speakers at $t$, and $M$ is the set of all masked time frames. 
Note that in our implementation, the output nodes for the primary and the secondary speakers are pre-determined (rather than solving the permutation using permutation invariant training \cite{yu2017permutation,wang2019speech,wang2020speaker}). This is because the utterance mixing algorithm guarantees the speech duration of the secondary speaker is substantially shorter than that of the primary speaker. 



\vspace{-.5em}
\subsection{Attention mask for streaming models}
\label{ssec:attention mask}
\vspace{-.5em}
Prior SSL models based on MSP objective \cite{hsu2021hubert,chen2022wavlm} utilized a standard self-attention in which every computation is executed by accessing the entire input sequence. This property restricts the SSL model to only offline scenarios. In this work, we adopt a masking strategy originally proposed for streaming TT~\cite{chen2021developing}, where a specially designed attention mask is applied to constrain the model to see only limited future information for each computation.

\begin{figure}[t]
    \centering
    \includegraphics[width=0.75\linewidth]{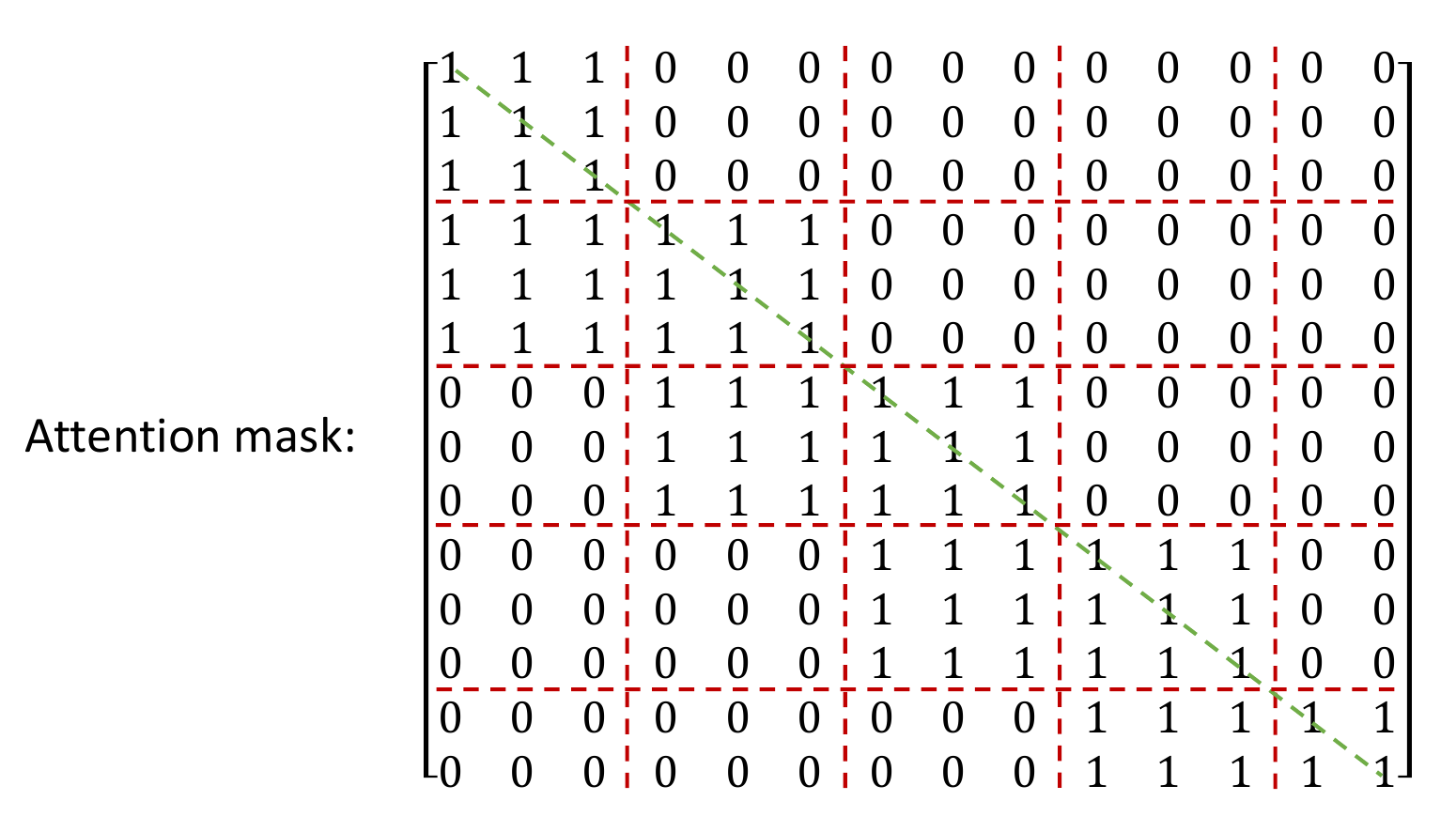}
    \vspace{-5mm}
    \caption{Attention mask matrix $\mathbf{S}$ for streaming models. If $\mathbf{S}[i, j] = 1$, the $j$th-frame input can be used for computing $i$th-frame output.}
    \label{fig:attention mask}
    \vspace{-3mm}
\end{figure}

The attention mask is defined as a $T\times T$ matrix $\mathbf{S}$, where $T$ is the embedding length, as exemplified in Fig.~\ref{fig:attention mask}. 
$\mathbf{S}[i, j] = 1$ means that the input at the $j$th frame can be used to compute the output at the $i$th frame. The matrix is segmented with fix-sized chunks for both vertical and horizontal directions (the chunk size is three in Fig. \ref{fig:attention mask}). Given the indices $\mathcal{I}_l$ corresponding for $l$-th chunk, the matrix $\mathbf{S}$ is defined such that $\mathbf{S}[i, j] = 1$ if ($i,j\in \mathcal{I}_l$ for any $l$) or ($i\in\mathcal{I}_l$ and $j\in\mathcal{I}_{l'}$ for $l-h<l'<l$), where $h$ is a hyper parameter to determine how far the history information can be accessed ($h=2$ in Fig.~\ref{fig:attention mask}).
With this masking strategy, the left receptive field (history) grows with the number of transformer layers while the right receptive field (future look-ahead) remains the same. The algorithm latency (or the duration of the future lookahead) of the model is determined by the chunk size.
In our work, we use the same masking matrix for the SSL-based pre-training and fine-tuning.

\vspace{-.5em}
\subsection{Data configurations for multi-talker ASR pre-training}
\label{ssec:other_explore}
\vspace{-.5em}
There are several data configurations that are especially important for multi-talker ASR pre-training. Firstly, data augmentation plays a crucial role to determine characteristics of extracted representation. In the case of WavLM, the training data was augmented such that a random noise was mixed to 10\% of the training samples while a random secondary speech was mixed to other 10\% of the training samples. Such a configuration effectively enforces the model to extract representation of the primary (i.e. dominant) speaker even from the overlapping speech. However, it has not yet been investigated if this configuration is optimum for the multi-talker tasks where the representations of all speakers are equally important. In this work, we thus explore to drastically increase the ratio of the utterance mixing for multi-talker modeling.






Secondary, we also explore several quantizers for generating the pseudo labels. The choice of quantizer is important in the MSP-based pre-training. For example, it is known that a quantizer with a higher correlation with phonemes generally benefits for ASR accuracy \cite{hsu2021hubert}. In this work, besides the clustering on FBANK or HuBERT embedding, we also investigate a phoneme-based quantizer proposed in~\cite{wang2022supervision}, in which a hybrid ASR system trained with a small amount of transcribed data is used for generating pseudo phoneme labels. We conduct various experiments to investigate the impact of quantizers in our experiment.

%
%
%

\begin{figure}[t]
    \centering
    \includegraphics[width=1.0\linewidth]{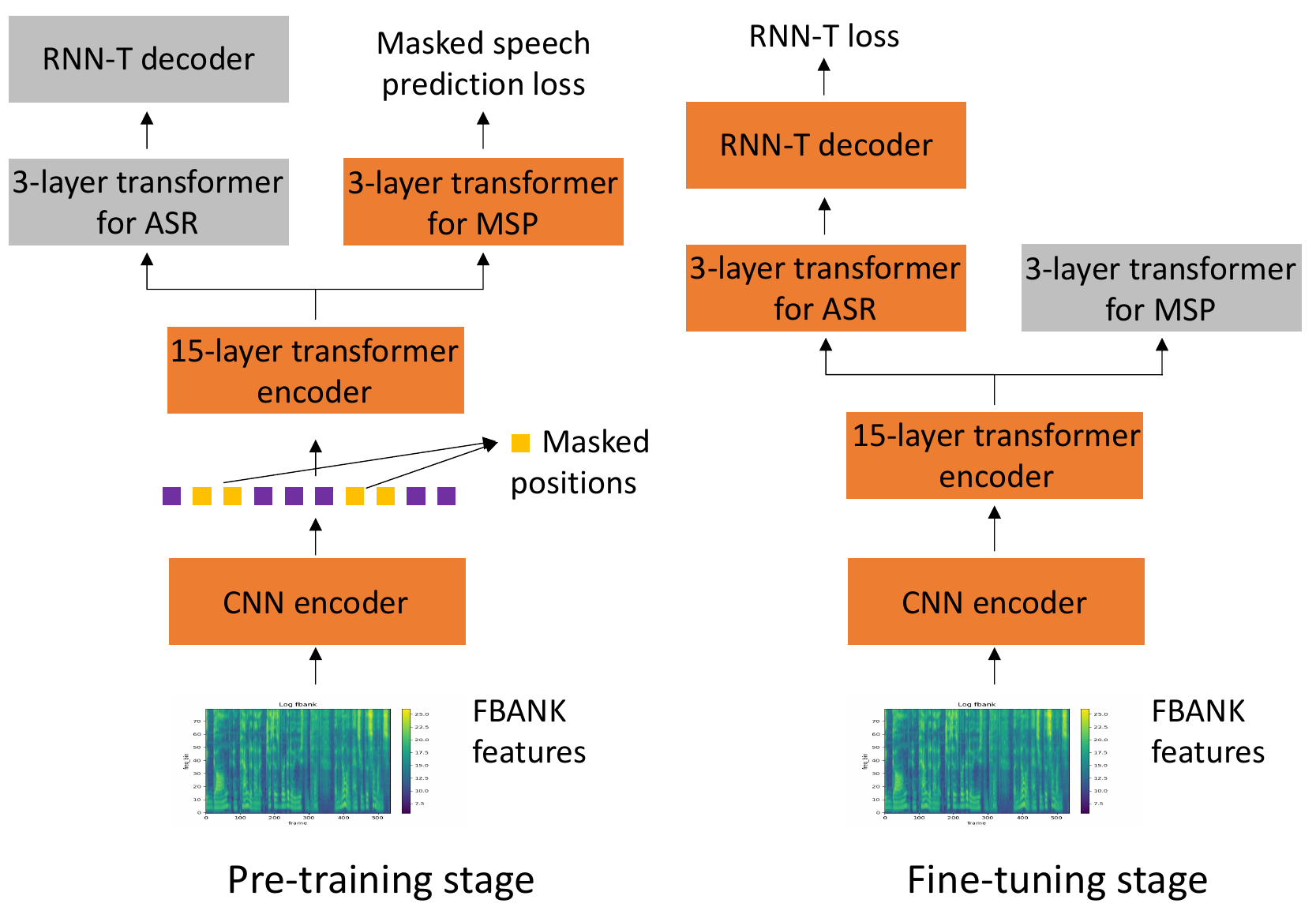}
    \vspace{-8mm}
    \caption{Pre-training pipeline for transformer transducer based t-SOT model. Orange blocks are updated while  grey blocks are frozen.}
    \label{fig:pipeline}
    \vspace{-5mm}
\end{figure}


\begin{table}[t]
\ra{0.9}
\tabcolsep = 1.5mm
    \centering
    \caption{WER (\%) on LibriSpeechMix for t-SOT TT-18 with different pre-training configurations. MSP stands for ``masked speech prediction". All models have 160 msec of algorithmic latency.}
    \begin{tabular}{cccccccc} \toprule
       \multicolumn{2}{c}{Pre-training} && \multicolumn{2}{c}{Dev WER (\%)} && \multicolumn{2}{c}{Test WER (\%)} \\ \cmidrule{1-2} \cmidrule{4-5} \cmidrule{7-8}
     Objective & Quantizer && 1spk & 2spk && 1spk & 2spk \\  \midrule
       - & -  && 15.42 & 39.12 && 15.69 & 39.52 \\ \hdashline[1pt/2pt]\hdashline[0pt/1pt]
       MSP          & FBANK   && 13.17 & 36.13 && 13.20 & 35.29 \\
       Bi-label MSP & FBANK   && 13.29 & 25.68 && 13.90 & 25.78 \\ \hdashline[1pt/2pt]\hdashline[0pt/1pt] 
       MSP          & HuBERT  && 10.77 & 17.24 && 11.30 & 17.25 \\
       Bi-label MSP & HuBERT  && 10.82 & 15.84 && 11.19 & 15.30 \\ \hdashline[1pt/2pt]\hdashline[0pt/1pt] 
       MSP          & Phoneme && 9.80 & 15.45 && 9.96 & 15.13 \\
       Bi-label MSP & Phoneme && \textbf{9.47} & \textbf{13.89} && \textbf{9.84} & \textbf{13.74} \\ \bottomrule
    \end{tabular}
    \label{tab:main}
    \vspace{-5mm}
\end{table}


\label{ssec:mixing strategy}
\begin{table}[t]
\ra{0.9}
\tabcolsep = 1.5mm
    \centering
    \caption{WER (\%) on LibriSpeechMix for t-SOT TT-18 with different data augmentation configuration for pre-training. We used the original MSP objective with the HuBERT quantizer. All models have 160 msec of algorithmic latency.}
    \begin{tabular}{ccccccccc} \toprule
       \multicolumn{3}{c}{Data augmentation} && \multicolumn{2}{c}{Dev WER (\%)} && \multicolumn{2}{c}{Test WER (\%)} \\ \cmidrule{1-3} \cmidrule{5-6} \cmidrule{8-9}
       No aug. & Noise & Speech && 1spk & 2spk && 1spk & 2spk \\ \midrule
        1.0 & -   & -   && 12.08 & 36.62 && 12.65 & 35.81 \\
        0.8 & 0.1 & 0.1 && 10.80 & 21.21 && 11.41 & 20.79 \\
        0.5 & - & 0.5 && \textbf{10.77} & \textbf{17.24} && \textbf{11.30} & \textbf{17.25} \\\bottomrule
    \end{tabular}
    \label{tab:mixing prob}
   \vspace{-5mm} 
\end{table}


\begin{table*}[t]
\ra{0.9}
    \centering
    \caption{WER (\%) on LibriSpeechMix test set for t-SOT TT-18 with different algorithmic latency $l$.}
    \begin{tabular}{cccccccccccccc} \toprule
        \multicolumn{2}{c}{Pre-training} && \multicolumn{2}{c}{$l=160$ msec} && \multicolumn{2}{c}{$l=640$ msec} && \multicolumn{2}{c}{$l=2560$ msec} && \multicolumn{2}{c}{$l=\infty$ (offline)} \\ \cmidrule{1-2} \cmidrule{4-5} \cmidrule{7-8} \cmidrule{10-11} \cmidrule{13-14}
        Objective    & Quantizer && 1spk & 2spk   && 1spk  & 2spk  && 1spk  & 2spk  && 1spk & 2spk \\ \midrule
        -            & -         && 15.69 & 39.52 && 13.71 & 34.71 && 12.21 & 29.56 && 11.00 & 24.03 \\
        Bi-label MSP & HuBERT    && 11.19 & 15.30 && 8.99  & 12.41 && 8.15  & 10.94 && 6.78 & 10.45 \\
        Bi-label MSP & Phoneme   && 9.84 & 13.74  && 8.39  & 11.24 && 7.20  & 9.55  && 6.46 & 8.51 \\ \bottomrule
    \end{tabular}
    \label{tab:latency}
    \vspace{-5mm}
\end{table*}

\vspace{-.5em}
\section{Experiments}
\label{sec:exp}
\vspace{-.5em}

\subsection{Experimental settings}
\label{ssec:setting}
\vspace{-.5em}

\subsubsection{Data}
\vspace{-.5em}
We evaluated our proposed framework on the LibriSpeechMix~\cite{kanda2020serialized} evaluation set. This dataset is simulated from LibriSpeech~\cite{panayotov2015librispeech} by randomly mixing utterances with random delays. In our experiments, we used the single-speaker and two-speaker-mixed evaluation sets. We adopt the same multi-talker ASR evaluation metric as~\cite{kanda2022streaming}. Specifically, we considered all possible speaker permutations between the hypotheses and references, and the permutation with the minimum number of errors was chosen to compute the word error rate (WER).

We trained our model with LibriSpeech 960h (LS-960), among which 100h of speech (train-clean-100, or LS-100) was used as labeled data and the other 860h (consisting of train-clean-360 and train-other-500) was considered unlabeled.


\vspace{-.5em}
\subsubsection{Model configuration}
\vspace{-.5em}


In our experiment, we used TT with the chunk-wise mask \cite{chen2021developing} as described in Section \ref{ssec:attention mask}, We used the same configuration as ``TT-18" in ~\cite{kanda2022streaming}. Specifically, the input to the network was 80-dim FBANK with a 10ms stride normalized by the mean and variance computed on the entire training data. The encoder consisted of 2 convolution layers which downsampled the acoustic features by a factor of 4, and 18 layers of transformers with relative positional encoding. Each transformer layer contained a 512-dim multi-head attention with 8 heads and a 2048-dim feed-forward layer. The prediction network of TT is a 2-layer LSTM network with 1024 hidden units. 

Our training pipeline is depicted in Fig.~\ref{fig:pipeline}. For pre-training, we optimized the bottom 15-layers of the encoder and three additional transformer layers with MSP or the proposed bi-label MSP objective on the LS-960. Once the model was pre-trained, we further fine-tuned the TT with RNN-T loss on simulated mixtures created from LS-100. The simulated mixtures for fine-tuning were generated on-the-fly such that the ratio between single-speaker and two-speaker-mixed samples is 50\%:50\%. The two-speaker-mixed samples were created by mixing two random utterances from LS-100. A random delay ranging from 0 to the duration of the first utterance was added to the second utterance to simulate partially overlapping speech. We also applied speed and volume perturbation to further increase the variety of the fine-tuning data.

We used the AdamW optimizer and a linear decay learning rate scheduler for both pre-training and fine-tuning. During pre-training, we trained the model on 16 NVIDIA V100 GPUs for 125k updates, with a batch size of 480s per GPU and a peak learning rate of 1.5e-3. During fine-tuning, we fine-tuned the model on 16 NVIDIA V100 GPUs for 35k updates, with a batch size of 60s per GPU and a peak learning rate of 3e-4.


As mentioned in section \ref{ssec:other_explore}, we experimented with three quantizers---FBANK, HuBERT, and phonemes. For the FBANK quantizer, the size of the codebook was set to $500$. For the HuBERT quantizer, we extracted hidden representations from the 9th layer of the HuBERT base model \footnote{The \href{https://dl.fbaipublicfiles.com/hubert/hubert_base_ls960.pt}{HuBERT base} model was pre-trained on LS-960.}, and clustered them into $500$ groups by using the K-means algorithm. For the phoneme quantizer, we first trained a hybrid ASR model on LS-100, and then decoded the LS-960 with it and a 3-gram language model. The hypothesis lattices were rescored and the phoneme labels were inferred from the 1-best path. There were $347$ distinct phonemes in total. 

\vspace{-.5em}
\subsection{Main results}
\label{ssec:main}
\vspace{-.5em}

Our main experimental results are shown in Table~\ref{tab:main}. As a baseline, we first trained a TT-18 model on the LS-100-based multi-talker fine-tuning data without any pre-training. The result is shown in the first row of Table~\ref{tab:main}. We found that, with such limited training data, the WERs were very bad for both the single-speaker and two-speaker-mixed evaluation sets. 

We then performed the pre-training based on the MSP objective with the FBANK quantizer, whose result is presented in the second row. During the pre-training, we applied utterance mixing with a 50\% probability. We first observed a large improvement in the single-speaker test set, where the WER was reduced by 15.9\% relatively (15.69\% to 13.20\%). On the other hand, we observed only 10.7\% of relative WER improvement (39.52\% to 35.29\%) for the two-speaker mixed test set. This result suggests that the MSP objective with utterance mixing was heavily biased towards the primary speaker, and it did not perform well for tasks that require modeling of all speakers.


After that, we replaced MSP with the proposed bi-label MSP objective, and the result is presented in the third row. Compared with the MSP result, we observed a significant WER reduction from 35.29\% to 25.78\% (26.9\% relative) on the two-speaker-mixed test set. As a side effect, there was a slight WER degradation on the single-speaker test set from 13.20\% to 13.90\%.


We finally evaluated the HuBERT-based quantizer and phoneme-based quantizer, whose results are listed for the last four rows. In this experiment, we again observed very consistent improvements in the two-speaker-mixed test sets by the proposed bi-label MSP objective. The best WER was achieved by using the phoneme-based quantizer. With the phoneme-based quantizer, the proposed bi-label MSP objective showed no side-effect even for the single-speaker test set while achieving remarkable improvement on the two-speaker-mixed test set.

\vspace{-.5em}
\subsection{Impact of the pre-training data configuration}
\label{ssec:distribution}
\vspace{-.5em}

We illustrate the impact of pre-training data configuration in Table~\ref{tab:mixing prob}. In this experiment, we used the conventional MSP with HuBERT quantizer for the pre-training. Here, the first row is the configuration without any data augmentation (used in the original HuBERT), the second row is the configuration used for WavLM, and the third row is the configuration used for our experiment described in the previous section. Pre-training without any data augmentation resulted in a very bad WER, especially for two-speaker-mixed evaluation sets (1st row). The ASR performance was significantly improved after adding a small ratio of noise and interference speech (2nd row). Finally, our configuration where utterance mixing is applied for 50\% of the pre-training data achieved the best WER for both single-speaker and two-speaker-mixed speech.

\vspace{-.5em}
\subsection{Latency variation}
\label{ssec:latency}

We also evaluated the model performance for different latency configurations, whose result is reported in Table~\ref{tab:latency}. We controlled the algorithmic latency by adjusting the chunk size in the self-attention mask as introduced in section~\ref{ssec:attention mask}. As shown in the first row, if not applying any pre-training, the WER for the two-speaker-mixed test set was over 20\%, even for the offline model. Our best-performing setup (bi-label MSP objective with the phoneme quantizer) dramatically improved the WER for all latency configurations. The 160ms latency model pre-trained with the bi-label MSP objective and the phoneme quantizer achieved better WER compared to the offline model without pre-training.



\vspace{-.5em}
\section{Conclusion}
\label{sec:conclusion}
\vspace{-.5em}


In this paper, we investigated SSL-based pre-training for streaming multi-talker ASR. We proposed the bi-label MSP objective which enforces the model to learn speech representations of all speakers rather than focusing on the primary speaker. We also explored several aspects of multi-talker ASR pre-training including the pre-training data configuration and the quantizer type. Our experimental results on LibriSpeechMix showed that our proposed SSL framework significantly reduced the WER on both non-overlapping and overlapping speech, especially with a prominent improvement on the overlapping speech.




\vfill\pagebreak

\bibliographystyle{IEEEbib}
\bibliography{strings,refs}

\end{document}